\documentclass[pra,onecolumn,a4paper,showpacs,showkeys,preprintnumbers,10pt,nofootinbib]{revtex4}
\usepackage{graphicx}
\usepackage{amsmath}
\usepackage{amsfonts}
\usepackage{amssymb}
\usepackage{bbm}
\usepackage{pst-plot}
\usepackage{pst-node}
\usepackage{pst-coil}
\usepackage{subfigure}
\newcommand{\tr}{{\rm{Tr}\,}}

\newcommand{\C}{\mathbb{C}}

\begin{document}

\title{Spin geometry of entangled qubits under bilocal decoherence modes}

\author{Katharina Durstberger}
\email{durstberger@ati.ac.at}
\affiliation{Atomic Institute of the Austrian Universities\\
Stadionallee 2, 1020 Vienna, Austria\\\vspace{1cm}}

\date{\today}

\begin{abstract}
The Lindblad generators of the master equation define which kind of
decoherence happens in an open quantum  system. We are working with a two qubit system and choose the generators to be projection operators on the eigenstates of the system and unitary bilocal rotations of them. The resulting decoherence modes are studied in detail. Besides the general solutions we investigate the special case of maximally entangled states -- the Bell singlet states. The results are depicted in the so-called spin geometry picture which allows to illustrate the evolution of the (nonlocal) correlations stored in a certain state. The question for which conditions the path traced out in the geometric picture depends only on the relative angle between the bilocal rotations is addressed.
\end{abstract}



\maketitle

\section{Introduction}

The theory of open quantum systems plays a major role in many applications of quantum physics
since perfect isolation of a quantum system is never possible. Because the environmental degrees
of freedom are not accessible the dynamics of open quantum systems are described by effective
dynamics: the quantum master equation \cite{BreuerPetruccione}. The notion of decoherence is introduced which describes the loss of quantum coherences in a system coupled to an external environment. This concept allows to understand the transition from a quantum to a classical world 
\cite{GiuliniJoosKieferKupschStamatescuZeh}. 

Geometric pictures of the quantum mechanical state space of a system are attracting and illustrative and provide deeper insight into unsolved problems. Therefore it is tempting to find geometric pictures for the whole state space of density matrices which is nearly impossible because even for the simplest case of a qubit the whole state space has $4$ real dimensions. 

By restricting to several properties one can find appealing pictures and visualizations of the state space (see for instance \cite{BengtssonZyczkowskiBook}). E.g., for pure two qubit states a Hopf map can be found which provides an entanglement sensitive stratification of the state space \cite{MosseriDandoloff}. The so-called spin geometric picture allows for another visualization of the two qubit system and was introduced and discussed by the Horodeckis 
\cite{Horodecki3_1997,Horodecki2_1996}, Vollbrecht and Werner \cite{VollbrechtWerner} and Bertlmann, Narnhofer and Thirring \cite{BertlmannNarnhoferThirring}.

The paper is organized in the following way. The next subsections give a brief introduction to the theory of decoherence with special emphasis on the theoretical formulation used later on in this paper and an introduction to the spin geometry picture where also the role of the singular value decomposition is emphasized. In Sect. \ref{sect.deco.modes} we introduce the decoherence modes under consideration which arise by local unitary rotations of the projection operators on the eigenstates of the system. We present the general solution of the time evolution of the local parameters $\vec m$ and $\vec n$ and the correlation matrix $c$ for three different types of decoherence, which we call decoherence modes. These general solutions are illuminated by considering special initial conditions: the Bell singlet state (Sect. \ref{sect:bell-state}). The evolution of the states under the investigated decoherence modes is visualized graphically within the spin geometry picture. In the last section \ref{sect:equiv} we show which restrictions on the initial correlation matrix $c$ of a state with local parameters equal to zero have to be satisfied such that the geometric pictures of mode B and C are equal.

\subsection{Decoherence}

A quantum system $S$ is coupled to the environment $E$ and the closed total system $S+E$ is
governed by unitary evolution given by the Hamilton operator
$H_{S+E}(t)=H\otimes\mathbbm1+\mathbbm1\otimes H_E+H_I$, where $H$ and $H_E$ are the free
Hamiltonians of the system and the environment, respectively, and $H_I$ denotes the interaction
Hamiltonian. Under several assumptions (see, e.g., \cite{BreuerPetruccione}) the reduced
non-unitary dynamics of the open system $S$ are given by the master equation
\begin{equation}
    \frac{\partial}{\partial t}\rho(t)=-i[H,\rho(t)]-\mathcal D(\rho(t))\;,
\end{equation}
with the dissipator \cite{Lindblad,GoriniKossakowskiSudarshan}
\begin{equation}\label{eq:dissipator,gen}
    \mathcal D(\rho)=\frac{1}{2}\sum_k \bigl(A_k^{\dag}A_k \rho +
    \rho A_k^{\dag}A_k-2A_k\rho A_k^{\dag}\bigr)\;.
\end{equation}
The general structure of the dissipator \eqref{eq:dissipator,gen} can be simplified by assuming the
a priori arbitrary Lindblad generators $A_k$ to be projection operators $A_k=\sqrt{\lambda_k}P_k$, with $P_k^2=P_k$ (see Ref. \cite{BertlmannGrimus2002}), which gives a simplified structure of the dissipator
\begin{equation}
    \mathcal D(\rho)=\frac{1}{2}\sum_k \lambda_k\bigl(P_k \rho +
    \rho P_k-2P_k\rho P_k\bigr)\;.
\end{equation}
The real and positive parameters $\lambda_k$ are called decoherence parameters and determine the strength of the interaction.

In the following treatments we work with a two qubit system where the Hilbert space is given by $\mathcal H=\C^2 \otimes \C^2$. We require the Lindblad generators $P_k$ to project onto one-dimensional subspaces and satisfy $\sum_{k=1}^4 P_k=\mathbbm{1}$. Furthermore we constrain to one dissipation parameter $\lambda$ which parameterizes the strength of the interaction and therefore of the decoherence \cite{BertlmannDurstbergerHasegawa2005}. Then the dissipator can be written in a very compact way
\begin{equation}\label{dissipator}
    \mathcal D(\rho)=\lambda\bigl(\rho- \sum_{k=1}^{4} P_k\rho P_k\bigr)\;,
\end{equation}
which is easier to deal with than the original one.

\subsection{Spin geometry}

The state space for a system consisting of two qubits has in general four complex or eight real dimensions. A general density matrix of such a system can be expressed in the following way 
\cite{Horodecki3_1997,Horodecki2_1996}
\begin{equation}\label{state-decomposition}
    \rho=\frac{1}{4}(\mathbbm{1}\otimes\mathbbm1 +\vec m\; \vec\sigma\otimes\mathbbm1 +\vec
    n\; \mathbbm1\otimes\vec\sigma+c_{ij}\; \sigma_i\otimes\sigma_j)\;,
\end{equation}
where $\{\mathbbm1\otimes\mathbbm1,\sigma_i\otimes\mathbbm1,\mathbbm1\otimes\sigma_j,
\sigma_i\otimes\sigma_j\}$ forms a basis in terms of product Pauli operators of $\mathcal
B(\mathcal H)=\mathcal B(\C^2\otimes\C^2)$, the algebra of bounded operators on the Hilbert space
$\mathcal H$.

The local parameters $\vec m,\vec n\in\mathbb{R}^3$ of the density matrix
\eqref{state-decomposition}, given by $\vec m=\tr(\vec\sigma\otimes\mathbbm{1}\;\rho)$ and $\vec
n=\tr(\mathbbm{1}\otimes\vec\sigma\;\rho)$, determine the reduced density matrices, e.g.,
$\rho_{1}=\tr_{2}\rho=\frac{1}{2}(\mathbbm1 + \vec m\cdot\vec\sigma)$, whereas the real $3\times3$
matrix $c=(c_{ij})$, where $c_{ij}=\tr(\sigma_i\otimes\sigma_j\;\rho)$, determines the nonlocal
correlations. The expectation value for a joint spin measurement in directions $\vec \alpha$
and $\vec\beta$, given by $E(\vec\alpha,\vec\beta)=\tr(\rho\;
\vec\alpha\cdot\vec\sigma\otimes\vec\beta\cdot\vec\sigma)
    =(\vec\alpha,c\; \vec\beta)$, is fully determined by the correlation matrix $c$.

A state is called separable if it can be written as a sum over product states, that means $c_{ij}=n_i\cdot m_j$. Nonseparable states are called entangled (see, e.g., \cite{BrussLeuchsBook}).
The amount of entanglement is invariant under unitary transformations of the form $U_1\otimes
U_2$. This provides us with an equivalence relation for states with equal properties concerning
separability and entanglement and we only have to choose a proper representative to reduce the number of parameters needed to describe the states.

The parameters $\vec m$, $\vec n$ and $c$ of $\rho$ transform under the action of the unitary transformation $U_1\otimes U_2$ as $\vec m'=O_1\vec m$, $\vec n'=O_2\vec n$ and $c'=O_1 c O_2^T$. The transformation $O_1$ ($O_2$) is related to $U_1$ ($U_2$) via the homomorphism connecting the groups $SU(2)$ and $SO(3)$: for every unitary transformation $U\in SU(2)$ there exists a unique rotation $O\in SO(3)$ such that $U \vec n\cdot\vec\sigma U^\dag=(O\vec n)\cdot\vec\sigma$.

It turns out that the orthogonal transformations can be chosen such that the matrix $c'$ is
diagonal. Thus it is sufficient to consider states as representatives where the
$c$-matrix is diagonal (singular value decomposition). The singular values, which are different
from the eigenvalues, are always real \footnote{Note that pure unitary dynamics do not affect the
degree of entanglement and thus do not appear in the singular value decomposition nor in the spin
geometry picture. For this reason it is legitimate to consider only the pure decoherence part of the equation, see Eq. \eqref{master}.}. They can be arranged to form a 3 dimensional vector $\vec c=(c_1,c_2,c_3)^T$ which we call correlation vector. The spin geometry picture consists of all possible correlation vectors $\vec c$.\\

\begin{figure}[htbp]
\centering
\includegraphics[width=4.5cm,keepaspectratio=true]{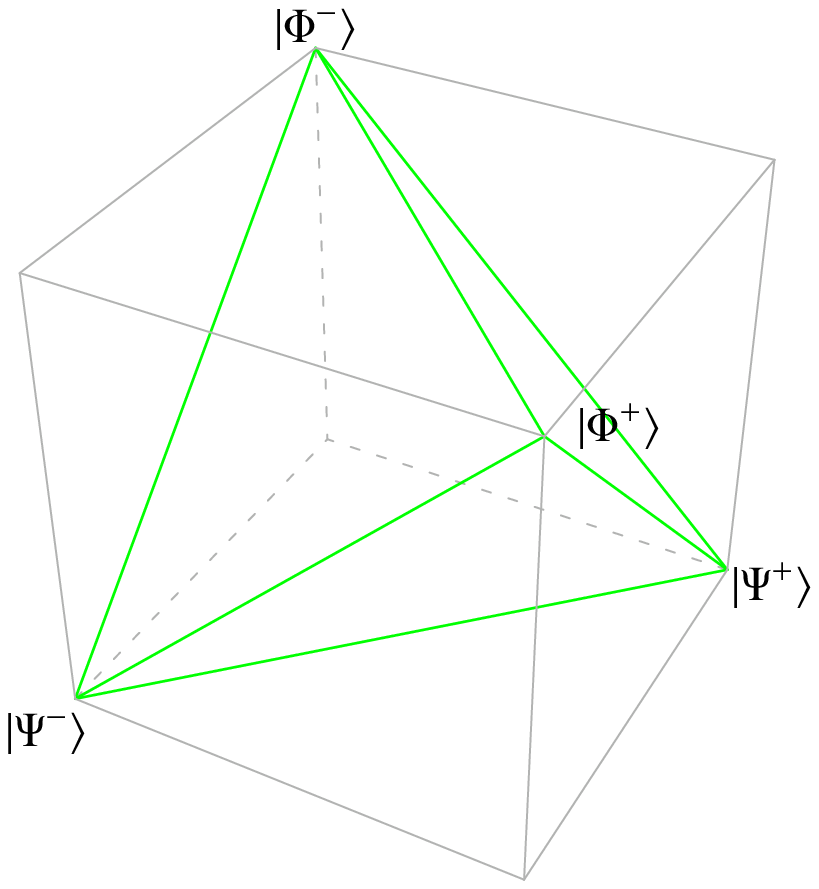}
\includegraphics[width=4.5cm,keepaspectratio=true]{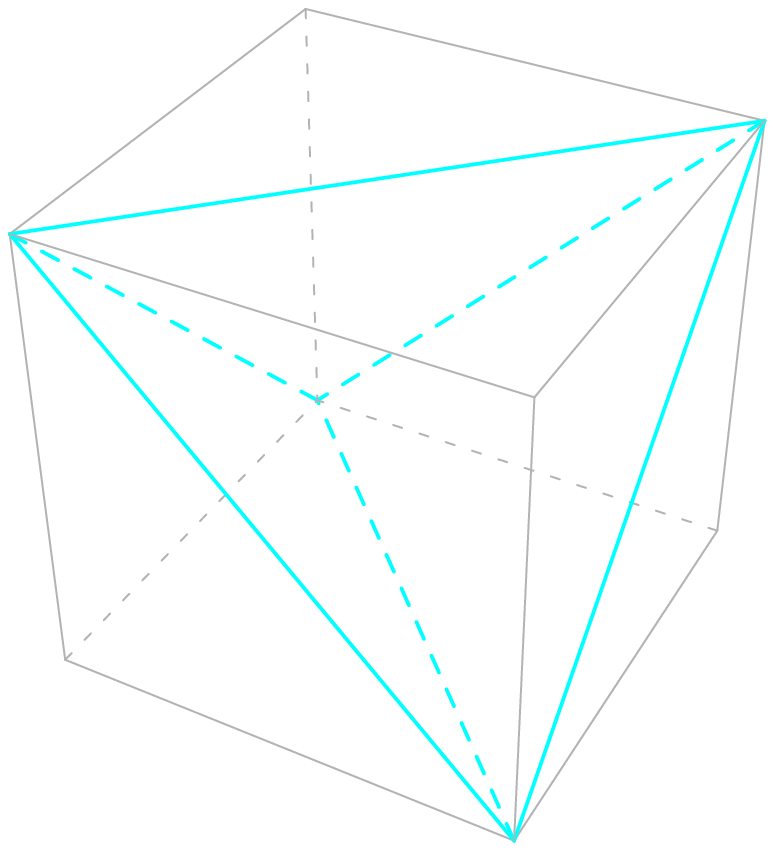}
\includegraphics[width=4.5cm,keepaspectratio=true]{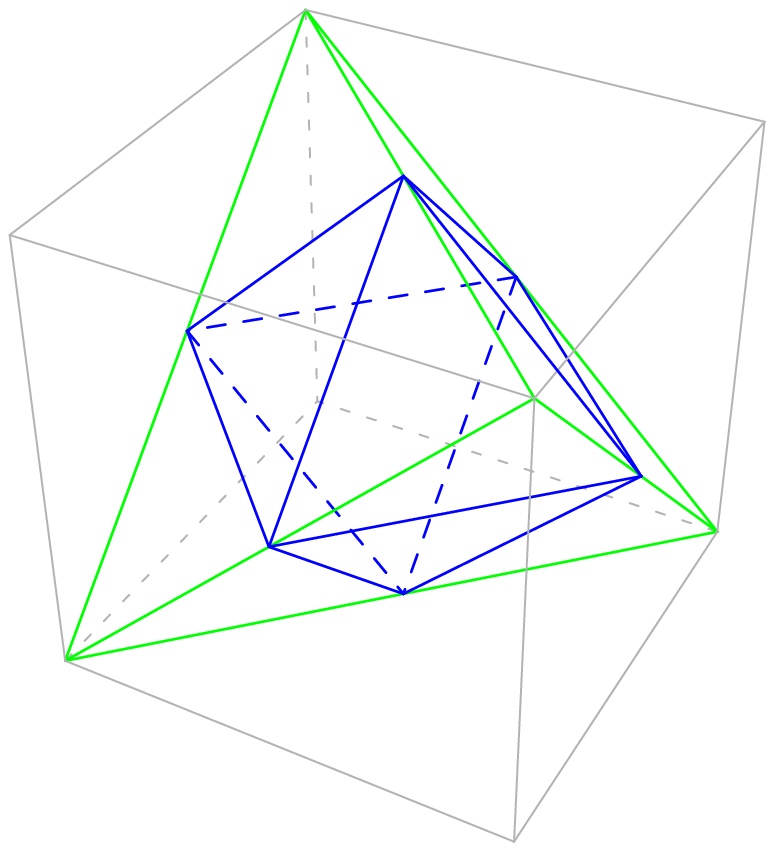}
  \caption{The tetrahedron of possible states, the inverted tetrahedron
  and the octahedron of separable states.}\label{fig.octahedron}
\end{figure}

\textbf{Example.}\\
The correlation vectors for the $4$ maximally entangled Bell states,
$\lvert\Psi^{\pm}\rangle=\frac{1}{\sqrt{2}}(\lvert01\rangle\pm\lvert10\rangle)$ with $\vec
c=(\pm1,\pm1,-1)^T$ and
$\lvert\Phi^{\pm}\rangle=\frac{1}{\sqrt{2}}(\lvert00\rangle\pm\lvert11\rangle)$ with $\vec
c=(\pm1,\mp1,+1)^T$, form the corners of a tetrahedron which characterizes the convex set of all
possible states. The partial transposition condition \cite{BrussLeuchsBook} leads to a
reflection of the tetrahedron and the set of separable state is given by the intersection of both
tetrahedrons which results in an octahedron (see Fig.~\ref{fig.octahedron} and Ref.~\cite{Horodecki2_1996}).

State with the same purity, measured by $\delta=\tr(\rho^2)$, correspond to a sphere of a radius proportional to $\delta$ centered at the origin  of the tetrahedron. There are 4 pure states with $\delta=1$ in the picture, the Bell states. States with equal entanglement, measured by the concurrence $C$ 
\cite{BrussLeuchsBook}, are found on planes with normal vector equal to $(-1,-1,-1)$ for instance. The only maximally entangled states in this picture are the Bell states. The separable states ($C=0$) lie at the border of the octahedron and inside. For a detailed analysis of these facts see Ref.~\cite{ZimannBuzek2005-2}.

\subsection{Singular value decomposition}

The singular value decomposition (SVD) is a generalization of the eigenvalue decomposition which can be used to diagonalize rectangular matrices (the eigen decomposition is defined only for square matrices). In analogy with the eigen decomposition, which decomposes a matrix into two simple matrices, the main idea of the SVD is to decompose a rectangular matrix $A$ into three simple matrices: two orthogonal matrices $P$ and $Q$ and one diagonal matrix $D$, given by
\begin{equation}\label{svd-trafo}
	A=P\;D\;Q^T\;,
\end{equation}
where $P^T P=\mathbbm1$ and $Q^T Q=\mathbbm1$. The diagonal entries of $D$ are called singular values. 

From the relations $A\,A^T=P\;D^2\;P^T$ and $A^TA=Q\;D^2\;Q^T$ it is obvious that that the eigenvalues of the matrices $A\,A^T$ and $A^TA$ correspond to the squares of the singular values of $A$ and the eigenstates of $A\,A^T$ and $A^TA$  form the columns of the matrices $P$ and $Q$. The columns of $P$ ($Q$) are called left (right) singular vectors of $A$.

\section{Decoherence modes}\label{sect.deco.modes}

The following investigations are concerned with decoherence effects. The aim
is to get information on the time evolution of the parameters $\vec m$, $\vec n$ and $c=(c_{ij})$,
introduced in Eq. \eqref{state-decomposition}, and to construct the spin geometry picture. Because the correlation vectors consists of the singular values of the correlation matrix where the dynamical evolution
has no impact on we can base our investigations on the master equation
\begin{equation}\label{master}
    \frac{\partial}{\partial t}\rho(t)=-\mathcal D(\rho(t))=\lambda\bigl(\rho- \sum_{k=1}^{4} P_k\rho P_k\bigr)\;,
\end{equation}
which describes only effects arising due to pure decoherence and includes no dynamical effects.

For a two qubit system $\{\lvert e_k\rangle\}_{k=1,\ldots, 4}$ denotes an eigenbasis of the Hamiltonian of the
undisturbed system, $H\lvert e_k\rangle=E_k\lvert e_k\rangle$.
In the following we chose several kinds of projection operators as Lindblad generators for
the dissipator, Eq. \eqref{master}. 

A one-dimensional (bilocal) projection operator $P_k$ on the eigenspace of the Hamiltonian $H$ can be written in terms of projections on the individual subspaces, e.g.,
\begin{equation}\label{proj.modeA}
P_k^A=\lvert e_k\rangle\langle e_k\rvert=P_k^{(1)}\otimes P_k^{(2)}\;,
\end{equation}
where $P_k^{(1)}=
\begin{cases}
\frac{1}{2}(\mathbbm1+\sigma_z)\quad \mbox{for}\; k=1,2\\
\frac{1}{2}(\mathbbm1-\sigma_z)\quad \mbox{for}\; k=3,4
\end{cases}$ and
$P_k^{(2)}=
\begin{cases}
\frac{1}{2}(\mathbbm1+\sigma_z)\quad \mbox{for}\; k=1,3\\
\frac{1}{2}(\mathbbm1-\sigma_z)\quad \mbox{for}\; k=2,4
\end{cases}$.

Decoherence modes where the Lindblad generators are projections of the form
\eqref{proj.modeA} are denoted as mode A. We consider also (bi-)unitary rotations of the
projection operators of the form
\begin{equation}\label{proj.modeB}
    P_k^B=(U_1\otimes \mathbbm1)P_k^A (U_1\otimes \mathbbm1)^\dag=
    U_1 P_k^{(1)}U_1^\dag\otimes P_k^{(2)}\;,
\end{equation}
which are called mode B, and more generally
\begin{equation}\label{proj.modeC}
    P_k^C=(U_1\otimes U_2)P_k^A (U_1\otimes U_2)^\dag=
    U_1 P_k^{(1)}U_1^\dag\otimes U_2 P_k^{(2)}U_2^\dag\;,
\end{equation}
which denote mode C. It is clear that mode A and B are just special cases of mode C.

A unitary rotation $U\in \rm{SU}(2)$ can be written in the following way
\begin{equation}
    U=e^{-i\frac{\alpha}{2}\vec
    a\cdot\vec\sigma}=\cos\frac{\alpha}{2}\mathbbm1-i\sin\frac{\alpha}{2}\vec a\cdot\vec\sigma=
    \begin{pmatrix}
        \cos\frac{\alpha}{2}-i a_z\sin\frac{\alpha}{2}\; & \;-(ia_x+a_y)\sin\frac{\alpha}{2}
        \\[5pt]
        -(ia_x-a_y)\sin\frac{\alpha}{2}\; & \;\cos\frac{\alpha}{2}+i a_z\sin\frac{\alpha}{2}  \\
    \end{pmatrix}\;,
\end{equation}
where $\alpha$ denotes the rotation angle and the unit vector $\vec a$ indicates the axis of
rotation. For convenience we chose $\vec a=(0,1,0)^T$. Thus the rotation matrix has the
simple form
\begin{equation}
    U(\alpha)=\begin{pmatrix}
        \cos\frac{\alpha}{2} & -\sin\frac{\alpha}{2} \\
        \sin\frac{\alpha}{2} & \cos\frac{\alpha}{2} \\
      \end{pmatrix}\;,
\end{equation}
and we set $U_1=U(\alpha)$ and $U_2=U(\beta)$ in Eqs. \eqref{proj.modeB} and \eqref{proj.modeC}.

\subsection{Mode A}

The first mode describes the simplest possible case. We choose the Lindblad generators according to Eq. \eqref{proj.modeA}. We have
to calculate the expression $P_k^A \rho P_k^A$ which gives the following, based on Eq.
\eqref{state-decomposition},
\begin{equation}
    P_k^A \rho P_k^A=\frac{1}{4}\biggl(P_k^A +
    \vec m\; (P_k^{(1)}\vec\sigma P_k^{(1)})\otimes P_k^{(2)} +
    \vec n\; P_k^{(1)} \otimes (P_k^{(2)}\vec\sigma P_k^{(2)})+
    c_{ij}\; (P_k^{(1)}\sigma_i P_k^{(1)})\otimes (P_k^{(2)}\sigma_j P_k^{(2)})\biggr)\,.
\end{equation}
After a short calculation we find that
\begin{equation}
\begin{split}
    P_k^{(1)}\sigma_i P_k^{(1)}=(\pm)^{k=1,2}_{k=3,4}\; \delta_{iz} P_k^{(1)}\;,\qquad
    P_k^{(2)}\sigma_i P_k^{(2)}=(\pm)^{k=1,3}_{k=2,4}\; \delta_{iz} P_k^{(2)}\;,\\
\end{split}
\end{equation}
where the signs are chosen accordingly and $\delta_{ij}$ denotes the Kronecker-delta, i.e.,
$\delta_{ij}=1$ for $i=j$ otherwise it is $0$. The sum over all k gives
\begin{equation}\label{sumA}
\begin{split}
    \sum_k P_k^A \rho P_k^A&=\frac{1}{4}\biggl(\mathbbm 1 +
    m_z\; (P_1^A+P_2^A-P_3^A-P_4^A) +
    n_z\; (P_1^A-P_2^A+P_3^A-P_4^A)+
    c_{zz}\; (P_1^A-P_2^A-P_3^A+P_4^A)\biggr)\\
    &=\frac{1}{4}\biggl(\mathbbm 1 +
    m_z\; \sigma_z\otimes\mathbbm1 +
    n_z\; \mathbbm1\otimes\sigma_z+
    c_{zz}\; \sigma_z\otimes\sigma_z\biggr)\;,
\end{split}
\end{equation}
where we have used the identities
\begin{equation}
\begin{split}
	P_1^A+P_2^A-P_3^A-P_4^A=\sigma_z\otimes\mathbbm1\;,\\
	P_1^A-P_2^A+P_3^A-P_4^A=\mathbbm1\otimes\sigma_z\;,\\
	P_1^A-P_2^A-P_3^A+P_4^A=\sigma_z\otimes\sigma_z\;.
\end{split}	
\end{equation}
With Eqs. \eqref{state-decomposition} and \eqref{sumA} we are ready to evaluate the master equation \eqref{master}.  A comparison of coefficients
leads to the following differential equations
\begin{equation}
\begin{split}
    \begin{pmatrix} \dot m_x \\ \dot m_y \\ \dot m_z \\ \end{pmatrix}
    = -\lambda\begin{pmatrix} m_x \\ m_y \\ 0 \\ \end{pmatrix}\;,\quad
    \begin{pmatrix} \dot n_x \\ \dot n_y \\ \dot n_z \\ \end{pmatrix}
    = -\lambda\begin{pmatrix} n_x \\ n_y \\ 0 \\ \end{pmatrix}\;,\quad
    \begin{pmatrix}
      \dot c_{xx} & \dot c_{xy} & \dot c_{xz} \\
      \dot c_{yx} & \dot c_{yy} & \dot c_{yz} \\
      \dot c_{zx} & \dot c_{zy} & \dot c_{zz} \\
    \end{pmatrix}=-\lambda
    \begin{pmatrix}
      c_{xx} & c_{xy} & c_{xz} \\
      c_{yx} & c_{yy} & c_{yz} \\
      c_{zx} & c_{zy} & 0 \\
    \end{pmatrix}\;,
\end{split}
\end{equation}
with the solutions
\begin{equation}\label{solutions.modeA}
\begin{split}
    m_z(t)=m_z(0),\quad n_z(t)=n_z(0),\quad c_{zz}(t)=c_{zz}(0)\\
    m_i(t)=e^{-\lambda t}m_i(0),\quad n_i(t)=e^{-\lambda t} n_i(0)\quad\mbox{for}\,i\neq z\\
    c_{ij}(t)=e^{-\lambda t} c_{ij}(0)\quad\mbox{for}\,i=j\neq z\;.
\end{split}
\end{equation}
This means that all elements are damped by $e^{-\lambda t}$ except the $z$-components of the local
parameters and the $zz$-component of the correlation matrix which are unaltered.

\subsection{Mode B}\label{sect.modeB}

The next stage is to consider rotations in one subspace, e.g., projection operators of the form
\eqref{proj.modeB}. Now the expression $P_k^B \rho P_k^B$ looks like
\begin{equation}\label{proj.sandwich.modeB}
\begin{split}
    P_k^B \rho P_k^B=\frac{1}{4}\biggl(P_k^B +
    &\vec m\;\; (U_1 P_k^{(1)}U_1^\dag \vec\sigma U_1 P_k^{(1)}U_1^\dag)\otimes P_k^{(2)}+
    \vec n\;\; (U_1 P_k^{(1)}U_1^\dag)  \otimes (P_k^{(2)}\vec\sigma P_k^{(2)})\\+
    &c_{ij}\;\; (U_1 P_k^{(1)}U_1^\dag \sigma_i U_1 P_k^{(1)}U_1^\dag )\otimes (P_k^{(2)}\sigma_j
    P_k^{(2)})\biggr)\;.
\end{split}
\end{equation}
With the expression
\begin{equation}\label{eq.u1}
\begin{split}
    U_1 P_k^{(1)}U_1^\dag \sigma_i U_1 P_k^{(1)}U_1^\dag =(\pm)^{k=1,2}_{k=3,4}\;
    (\delta_{ix} \sin\alpha +\delta_{iz}\cos\alpha) U_1 P_k^{(1)}U_1^\dag\;,
\end{split}
\end{equation}
we get for the sum
\begin{equation}
\begin{split}
    \sum_k P_k^B \rho P_k^B=\frac{1}{4}\biggl(\mathbbm 1 &+
        (m_x \sin\alpha+m_z\cos\alpha)\; (\sin\alpha\, \sigma_x\otimes\mathbbm1+\cos\alpha\, \sigma_z\otimes\mathbbm1) +
        n_z\; \mathbbm1\otimes\sigma_z\\
        &+(c_{xz}\sin\alpha + c_{zz}\cos\alpha)\;
        (\sin\alpha \,\sigma_x\otimes\sigma_z+\cos\alpha\, \sigma_z\otimes\sigma_z)\biggr)\;.
\end{split}
\end{equation}
The differential equations for the parameters of the density matrix are given by
\begin{equation}
\begin{split}
    \begin{pmatrix} \dot m_x \\ \dot m_y \\ \dot m_z \\ \end{pmatrix}
    = -\lambda\begin{pmatrix} m_x-m_x \sin^2\alpha-m_z\sin\alpha\cos\alpha \\ m_y \\
    m_z-m_x \sin\alpha\cos\alpha-m_z\cos^2\alpha \\ \end{pmatrix}\;,\quad
    \begin{pmatrix} \dot n_x \\ \dot n_y \\ \dot n_z \\ \end{pmatrix}
    = -\lambda\begin{pmatrix} n_x \\ n_y \\ 0 \\ \end{pmatrix}\,,\\
    \begin{pmatrix}
      \dot c_{xx} & \dot c_{xy} & \dot c_{xz} \\
      \dot c_{yx} & \dot c_{yy} & \dot c_{yz} \\
      \dot c_{zx} & \dot c_{zy} & \dot c_{zz} \\
    \end{pmatrix}=-\lambda
    \begin{pmatrix}
      c_{xx}\; & c_{xy}\; & c_{xz}-c_{xz}\sin^2\alpha-c_{zz}\sin\alpha\cos\alpha \\[5pt]
      c_{yx}\; & c_{yy}\; & c_{yz} \\[5pt]
      c_{zx}\; & c_{zy}\; & c_{zz}-c_{xz}\sin\alpha\cos\alpha-c_{zz}\cos^2\alpha \\[5pt]
    \end{pmatrix}\;.
\end{split}
\end{equation}
Apart from the solutions already given in Eq. \eqref{solutions.modeA} we get the following
solutions for the remaining components
\begin{equation}\label{sol.modeB-m}
\begin{split}
    m_{x}(t)=(\sin^2\alpha+e^{-\lambda t}\cos^2\alpha)m_{x}(0) +(1-e^{-\lambda t})\sin\alpha\cos\alpha\, m_{z}(0)   \;,\\
    m_{z}(t)=(1-e^{-\lambda t})\sin\alpha\cos\alpha\,m_{x}(0) + (e^{-\lambda t}\sin^2\alpha+\cos^2\alpha)m_{z}(0) \;,
\end{split}
\end{equation}
\begin{equation}\label{sol.modeB-c}
\begin{split}
    c_{xz}(t)= (\sin^2\alpha+e^{-\lambda t}\cos^2\alpha)c_{xz}(0) +(1-e^{-\lambda t})\sin\alpha\cos\alpha\, c_{zz}(0)\;,\\
    c_{zz}(t)= (1-e^{-\lambda t})\sin\alpha\cos\alpha\,c_{xz}(0) + (e^{-\lambda t}\sin^2\alpha+\cos^2\alpha)c_{zz}(0)\;.
\end{split}
\end{equation}

This means the rotation of the projection operators results in a coupling of different components, in this case between $m_x$ and $m_z$ and $c_{xz}$ and $c_{zz}$.

This mode has been discussed in detail in Ref.\cite{BertlmannDurstbergerHasegawa2005}.

\subsection{Mode C}

Independent rotations in both subspaces, where the projection operators are given by Eq. \eqref{proj.modeC}, results in an expression like
\begin{equation}
\begin{split}
    P_k^C \rho P_k^C=\frac{1}{4}\biggl(P_k^C +
    &\,\vec m\;\; (U_1 P_k^{(1)}U_1^\dag \vec\sigma U_1 P_k^{(1)}U_1^\dag)\otimes (U_2 P_k^{(2)}U_2^\dag)+
    \vec n\;\; (U_1 P_k^{(1)}U_1^\dag)  \otimes (U_2 P_k^{(2)}U_2^\dag \vec\sigma U_2 P_k^{(2)}U_2^\dag)\\+
    &\,c_{ij}\;\; (U_1 P_k^{(1)}U_1^\dag \sigma_i U_1 P_k^{(1)}U_1^\dag )\otimes
                (U_2 P_k^{(2)}U_2^\dag \sigma_j U_2 P_k^{(2)}U_2^\dag)\biggr)\;,
\end{split}
\end{equation}
where $U_1 P_k^{(1)}U_1^\dag \sigma_i U_1 P_k^{(1)}U_1^\dag$ is given in Eq. \eqref{eq.u1} and
\begin{equation}
\begin{split}
    U_2 P_k^{(2)}U_2^\dag \sigma_i U_2 P_k^{(2)}U_2^\dag =(\pm)^{k=1,3}_{k=2,4}\;
    (\delta_{ix} \sin\beta +\delta_{iz}\cos\beta) U_2 P_k^{(1)}U_2^\dag\;.
\end{split}
\end{equation}
Calculating the sum over all 4 terms gives
\begin{equation}
\begin{split}
    \sum_k P_k^C \rho P_k^C=\frac{1}{4}\biggl(\mathbbm 1 &+
        \bar m\; (\sin\alpha\, \sigma_x\otimes\mathbbm1+\cos\alpha\, \sigma_z\otimes\mathbbm1) +
        \bar n\; (\sin\beta\, \mathbbm1\otimes\sigma_x+\cos\beta\, \mathbbm1\otimes\sigma_z)\\
        &+\bar c\;
        (\sin\alpha\sin\beta \,\sigma_x\otimes\sigma_x+\sin\alpha\cos\beta
        \,\sigma_x\otimes\sigma_z
        +\cos\alpha\sin\beta\, \sigma_z\otimes\sigma_x+\cos\alpha\cos\beta\,
        \sigma_z\otimes\sigma_z)\biggr)\;,
\end{split}
\end{equation}
where
\begin{equation}
\begin{split}
    \bar m&=m_x \sin\alpha+m_z\cos\alpha\;,\hspace{1cm}
    \bar n=n_x \sin\beta+n_z\cos\beta\;,\\
    \bar c&=c_{xx}\sin\alpha\sin\beta +c_{xz}\sin\alpha\cos\beta +
        c_{zx}\cos\alpha\sin\beta+ c_{zz}\cos\alpha\cos\beta\;.
\end{split}
\end{equation}
The differential equations for the parameters of the density matrix are given by
\begin{equation}
\begin{split}
    \begin{pmatrix} \dot m_x \\ \dot m_y \\ \dot m_z \\ \end{pmatrix}
    = -\lambda&\begin{pmatrix} m_x-m_x \sin^2\alpha-m_z\sin\alpha\cos\alpha \\ m_y \\
    m_z-m_x \sin\alpha\cos\alpha-m_z\cos^2\alpha \\ \end{pmatrix}\,,\quad
    \begin{pmatrix} \dot n_x \\ \dot n_y \\ \dot n_z \\ \end{pmatrix}
    = -\lambda\begin{pmatrix} n_x-n_x \sin^2\beta-n_z\sin\beta\cos\beta \\ n_y \\
    n_z-n_x \sin\beta\cos\beta-n_z\cos^2\beta \\ \end{pmatrix}\,,\\
    &\begin{pmatrix}
      \dot c_{xx} & \dot c_{xy} & \dot c_{xz} \\
      \dot c_{yx} & \dot c_{yy} & \dot c_{yz} \\
      \dot c_{zx} & \dot c_{zy} & \dot c_{zz} \\
    \end{pmatrix}=-\lambda
    \begin{pmatrix}
      c_{xx}-\bar c\, \sin\alpha\sin\beta \; & c_{xy}\; & c_{xz}-\bar c\,\sin\alpha\cos\beta  \\[5pt]
      c_{yx}\; & c_{yy}\; & c_{yz} \\[5pt]
      c_{zx}-\bar c\,\cos\alpha\sin\beta \; & c_{zy}\; & c_{zz}-\bar c\,\cos\alpha\cos\beta \\[5pt]
    \end{pmatrix}\;.
\end{split}
\end{equation}

The solutions for $\vec m(t)$ and $\vec n(t)$ have the same structure, given by Eq.
\eqref{sol.modeB-m}. The solutions for the modified components of the $c$-matrix are the following
\begin{equation}\label{eq:sol.modeC}
\begin{split}
    c_{xx}(t)=e^{-\lambda t}c_{xx}(0)+ (1-e^{-\lambda t})
    \biggl(&\sin^2\alpha\sin^2\beta\; c_{xx}(0)+\sin^2\alpha\sin\beta\cos\beta\; c_{xz}(0)\\
    &+\sin\alpha\cos\alpha\sin^2\beta\; c_{zx}(0)+\sin\alpha\cos\alpha\sin\beta\cos\beta \;c_{zz}(0)\biggr)\,,\\
    c_{xz}(t)=e^{-\lambda t}c_{xz}(0)+ (1-e^{-\lambda t})
    \biggl(&\sin^2\alpha\sin\beta\cos\beta\; c_{xx}(0)+\sin^2\alpha\cos^2\beta\; c_{xz}(0)\\
    &+\sin\alpha\cos\alpha\sin\beta\cos\beta \;c_{zx}(0)+\sin\alpha\cos\alpha\cos^2\beta\; c_{zz}(0)\biggr)\;,\\
    c_{zx}(t)=e^{-\lambda t}c_{zx}(0)+ (1-e^{-\lambda t})
    \biggl(&\sin\alpha\cos\alpha\sin^2\beta\; c_{xx}(0)+\sin\alpha\cos\alpha\sin\beta\cos\beta\; c_{xz}(0)\\
    &+\cos^2\alpha\sin^2\beta\; c_{zx}(0)+\cos^2\alpha\sin\beta\cos\beta\; c_{zz}(0)\biggr)\;,\\
    c_{zz}(t)=e^{-\lambda t}c_{zz}(0)+ (1-e^{-\lambda t})
    \biggl(&\sin\alpha\cos\alpha\sin\beta\cos\beta\; c_{xx}(0)+\sin\alpha\cos\alpha\cos^2\beta \;c_{xz}(0)\\
    &+\cos^2\alpha\sin\beta\cos\beta\; c_{zx}(0)+\cos^2\alpha\cos^2\beta\; c_{zz}(0)\biggr)\;.\\
\end{split}
\end{equation}

Now the coupling in the $c$-matrix is extended to include $c_{xx}$, $c_{xz}$, $c_{zx}$ and $c_{zz}$ components. Therefore the solutions for this mode are more demanding. 

\subsection{Comments}

We see from the structure of the solutions for mode C, Eq. \eqref{eq:sol.modeC}, that the rotation axis of the unitary rotation has an influence on the resulting local parameters and the correlation matrix. The different modes correspond to the coupling of the system with different environments. The coupling strength parameterized by $\lambda$ is always the same but the effects on the state of the system are different.  

To calculate the correlation vector for all these modes is in general not very easy because one gets quite big terms. Therefore we postpone this investigation to the next section where we calculate the vectors for the special case of the Bell singlet state.

An interesting point is what happens with the correlation matrix for $t\rightarrow\infty$. As can be easily checked the asymptotic correlation matrix is given by
\begin{equation}\label{eq.limit-states}
	c^C_\infty=w\begin{pmatrix}
		 \sin\alpha \sin\beta & 0 &\sin\alpha \cos\beta  \\
		 0 & 0 & 0 \\
 		\cos\alpha \sin\beta & 0 & \cos\alpha \cos\beta
	\end{pmatrix}\;,
\end{equation}
where $w=\sin\alpha \bigl(c_{xz}(0) \cos\beta+c_{xx}(0) \sin\beta\bigr)+\cos\alpha\bigl(c_{zz}(0) \cos\beta+c_{zx}(0) \sin\beta\bigr)$. The correlation vector for the asymptotic matrix has the simple structure of 
$\vec c_\infty=(0, 0 , w)^T$. This shows that all states independent of their initial correlations end up on a line connecting the origin of the spin geometry picture (the maximally mixed state) with a point representing an equal mixture of two Bell states given by the corners of the octahedron (see Fig. \ref{fig.limit}). Note that the ordering of the singular values does not matter due to the high symmetry of the picture.

\section{Example: Bell state}\label{sect:bell-state}

To get a better feeling for the decoherence modes we consider the example of the maximally
entangled Bell singlet state $\lvert\Psi^-\rangle=\frac{1}{\sqrt{2}}(\lvert e_2\rangle-\lvert e_3\rangle)=\frac{1}{\sqrt{2}}(0,1,-1,0)^T$ where the initial density matrix is given by $\rho(0)=\lvert\Psi^-\rangle\langle\Psi^-\rvert$.
The local parameters vanish $\vec m=\vec n=0$ and the correlation matrix is diagonal $c_{BS}(0)={\rm
diag}(-1,-1,-1)$. In the following we consider the behavior of the correlation matrix
because the local parameters do not change for this state. 

\subsection{Mode A}

For decoherence mode A the correlation matrix remains diagonal but is affected by the
decoherence like $c^A_{BS}={\rm diag}(-e^{-\lambda t},-e^{-\lambda t},-1)$. Consequently the correlation vector
for mode A is given by $\vec c^A_{BS}=(-e^{-\lambda t},  -e^{-\lambda t} , -1 )^T$.

The correlation vector $\vec c^A_{BS}$ for fixed $\lambda$ is plotted in Fig.\ref{fig.modeA} with respect
to varying $t$. We start in the corner indicated by $\lvert\Psi^-\rangle$ and for
$t\rightarrow\infty$ approach the point bisecting the line which connects the projectors of the
states $\lvert\Psi^-\rangle$ and $\lvert\Psi^+\rangle$. This asymptotic state is mixed but not
maximally mixed and lies at the border of separability which is given by the blue octahedron.

\subsection{Mode B}\label{example1,modeB}

Mode B, where the projector in one subspace is rotated unitarily, leads to a
correlation matrix given by
\begin{equation}
    c^B_{BS}=\\
    \begin{pmatrix}
      -e^{-\lambda t} & 0 & -(1-e^{-\lambda t})\sin\alpha\cos\alpha \\
      0 & -e^{-\lambda t} & 0 \\
      0 & 0 & -e^{-\lambda t}-(1-e^{-\lambda t})\cos^2\alpha\\
    \end{pmatrix}\;.
\end{equation}
The emergence of the $xz$-component is due to the particular coupling of the differential equations, Eq. \eqref{sol.modeB-c}. The singular values \footnote{Note, that the SVD does not specify a certain sign, which can be determined, e.g., from the eigenvalue decomposition, nor a certain attribution to the coordinate axes, but note the high symmetry of the tetrahedron.} 
of this matrix are computed to be
\begin{equation}\label{sing.values-modeB}
\begin{split}
    c^B_1&=e^{-\lambda t}\;,\\
    c^B_{2,3}&=\frac{e^{-\lambda t}}{2}\sqrt{3+2e^{2\lambda t}\cos^2\alpha-\cos(2\alpha)\mp
        \sqrt{2}(e^{\lambda t}-1)\cos\alpha\sqrt{5-3\cos(2\alpha)
        +2e^{\lambda t}(2+e^{\lambda t})\cos^2\alpha}}\;.
\end{split}
\end{equation}
which gives for the correlation vector $\vec c^B_{BS}=(-c^B_1,-c^B_2,-c^B_3)^T$. 

The expansions of the last two singular values up to second order in $\alpha$,
\begin{equation}
\begin{split}
    c^B_2=e^{-\lambda t}+\frac{e^{-\lambda t}}{2}\frac{1-e^{\lambda t}}{1+e^{\lambda
    t}}\;\alpha^2\;,\qquad
    c^B_3=1+\frac{e^{-\lambda t}}{2}\frac{1-e^{\lambda t}}{1+e^{\lambda t}}\,(2+e^{\lambda t})\;\alpha^2\;,
\end{split}
\end{equation}
show that to first order in $\alpha$ the singular value $c_2$ is well approximated by $e^{-\lambda
t}$ and $c_3$ is constant. The second order contributions affect $c_3$ much more than $c_2$. In fact the deviations of $c_2$ from the exponential function are very small (a few percent) and the largest deviation arises for $\alpha=\frac{3\pi}{8}$. 

The decoherence paths for mode B form a plane which has a little bulge due to the slight deviation of $c^B_2$ from the exponential function. This deviation from the plane formed by equal $c^B_1$ and $c^B_2$ coefficients can be understood by the fact that the SVD is an asymmetric transformation and although applied to square matrices the matrices $P$ and $Q$, Eq.\eqref{svd-trafo}, are different. 
A comparison with the eigenvalues of $c^B_{BS}$, given by 
$\lambda^B_1=\lambda^B_2=-e^{-\lambda t}$ and
$\lambda^B_3=-\cos^2\alpha-e^{-\lambda t}\sin^2\alpha$, reveals that only the $z$-component depends on the angle $\alpha$. 

\begin{figure}[htbp]%
\centering
\subfigure[]{\includegraphics[width=4.5cm,keepaspectratio=true]{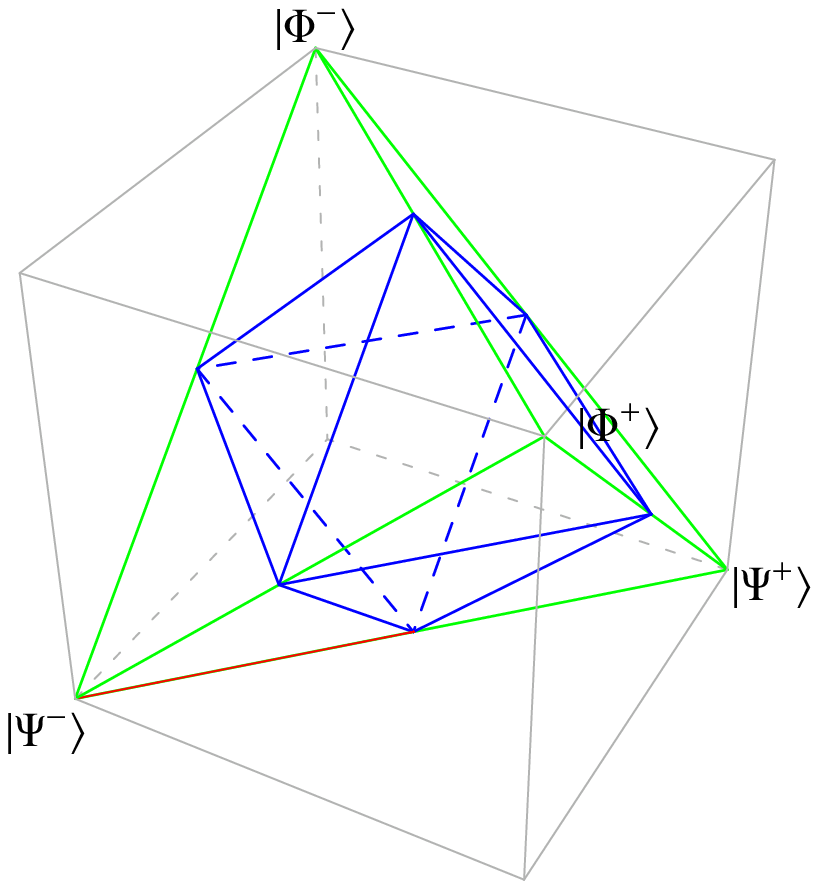}\label{fig.modeA}}\qquad
\subfigure[]{\includegraphics[width=4.5cm,keepaspectratio=true]{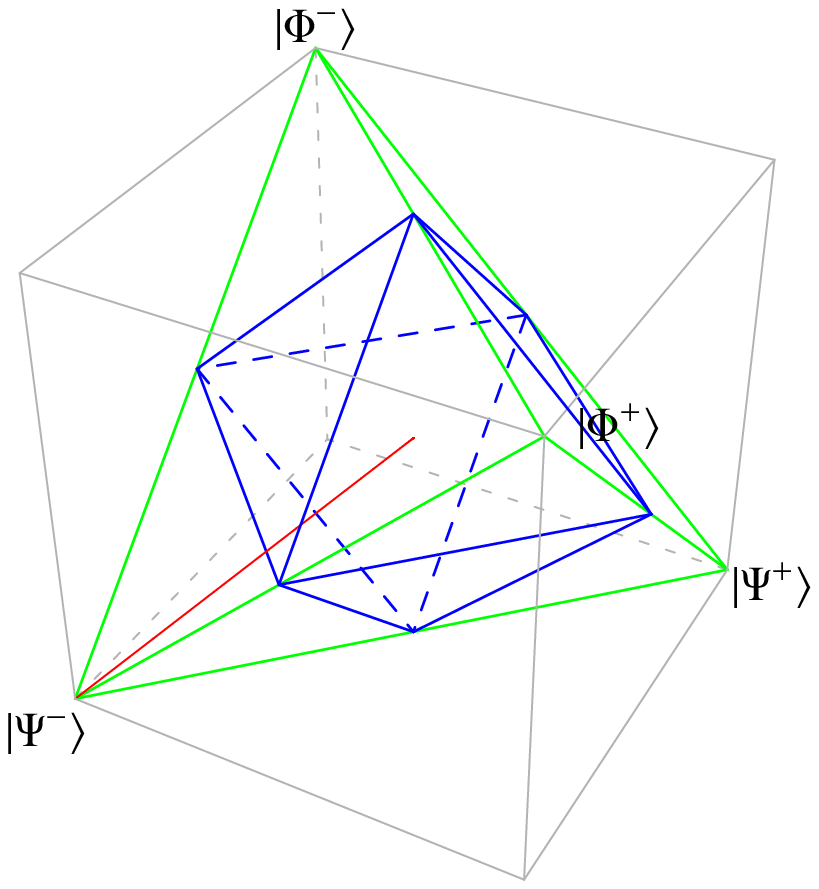}\label{fig.modeB-1}}\qquad
\subfigure[]{\includegraphics[width=4.5cm,keepaspectratio=true]{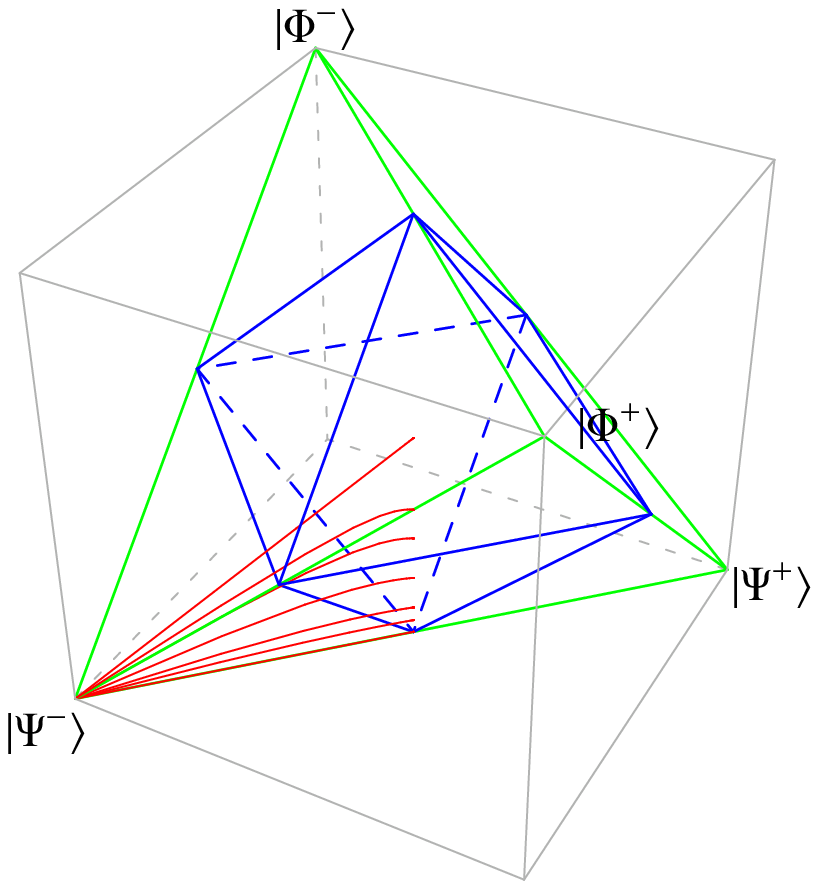}\label{fig.modeB-2}}%
\caption{The red lines represent possible decoherence paths of mode A and B for the Bell singlet state. Figure \subref{fig.modeA} shows the path traced out by mode A, \subref{fig.modeB-1} depicts
mode B for $\alpha=\frac{\pi}{2}$ (the line of the Werner state) and \subref{fig.modeB-2} visualizes mode B
  for different $\alpha$-values in steps of $\frac{\pi}{10}$.} \label{fig.modeB}
\end{figure}

We can distinguish two special cases. The correlation vector for $\alpha=0$ is given by 
$\vec c=( -e^{-\lambda t}, -e^{-\lambda t}, -1)^T$ which corresponds exactly to mode A (see Fig. 
\ref{fig.modeA}). For $\alpha=\frac{\pi}{2}$ the correlation vector, given by $\vec c=( -e^{-\lambda t}, -e^{-\lambda t}, -e^{-\lambda t})^T$, is shown in Fig. \ref{fig.modeB-1}. The state approaches the totally mixed state sitting at the origin of the coordinate system. Thereby it reaches the border of separability at $\lambda t=\ln3$ (cf. with Ref. \cite{BertlmannDurstbergerHasegawa2005}). We recover the phenomenon of ``entanglement sudden death'' introduced by Yu and Eberly \cite{YuEberly2006}. Note that the Werner state 
\cite{Werner1989}, which interpolates between a maximally entangled state and the maximally mixed state, shows the same behaviour in the spin geometry picture.

In Fig. \ref{fig.modeB-2} the correlation vectors are shown with respect to fixed parameter $\lambda$ and evolving time $t$ for different values of $\alpha$. The time the border of separability is reached varies
with respect to $\alpha$ and the extremal cases are $\alpha=\frac{\pi}{2}$ where the border is
reached after the shortest time and $\alpha=0$ where it is reached asymptotically at infinity.

\subsection{Mode C}

The most general case, mode C, which is characterized by two bilocal unitary rotations of the Lindblad generators, exhibits a time evolved correlation matrix for the Bell singlet state given by
\begin{equation}
    c^C_{BS}=
    \begin{pmatrix}
      -e^{-\lambda t}- (1-e^{-\lambda t}) \cos (\alpha -\beta ) \sin\alpha\sin\beta & 0 
      & - \left(1-e^{-\lambda t}\right) \cos (\alpha -\beta ) \sin\alpha\cos\beta\\
      0 & -e^{-\lambda t} & 0 \\
      -\left(1-e^{-\lambda t}\right) \cos (\alpha -\beta )\cos\alpha\sin\beta
   & 0 & -e^{-\lambda t} -(1-e^{-\lambda t}) \cos (\alpha -\beta ) \cos\alpha\cos\beta\\
    \end{pmatrix}\;,
\end{equation}
We calculate the singular values,
\begin{equation}
\begin{split}
    c^C_1&=e^{-\lambda t}\;,\\
    c^C_{2,3}&=\frac{1}{2}e^{-\lambda t}
    \sqrt{3+2e^{2\lambda t}\cos^2\Delta-\cos(2\Delta)\mp
        \sqrt{2}(e^{\lambda t}-1)\cos\Delta\sqrt{5-3\cos(2\Delta)
        +2e^{\lambda t}(2+e^{\lambda t})\cos^2\Delta}}\;,
\end{split}
\end{equation}
where $\Delta=\alpha-\beta$ and realize that they have the same structure as 
Eq. \eqref{sing.values-modeB} with a symmetry in $\alpha$ and $\beta$. They depend only on the relative angle difference $\lvert\Delta\rvert$ and consequently the problem reduces to mode B, already discussed in section \ref{example1,modeB}.

Just for completeness, the eigenvalues of the correlation matrix for mode C are given by 
$\lambda^C_{1}=\lambda^C_{2}=-e^{-\lambda t}$ and $\lambda^C_3=-\cos^2\Delta-e^{-\lambda t}\sin^2\Delta$.

\section{Equivalence of mode B and mode C in the geometric picture}\label{sect:equiv}

The results of the last section suggest to investigate the point which initial states result in an equivalence of mode B and C in the spin geometry picture. This means that for some initial conditions the singular values (and also the eigenvalues) of the correlations matrix for mode C depend only on the angle difference $\Delta=\alpha-\beta$ and have the same structure as those for mode B. We state a proposition which is a necessary condition for the initial correlation matrix that mode C depends only on the angle difference and is equivalent to mode B. We do not know if the condition is also sufficient.\\

\begin{figure}[htbp]%
\centering
\subfigure[]{\includegraphics[width=4.5cm,keepaspectratio=true]{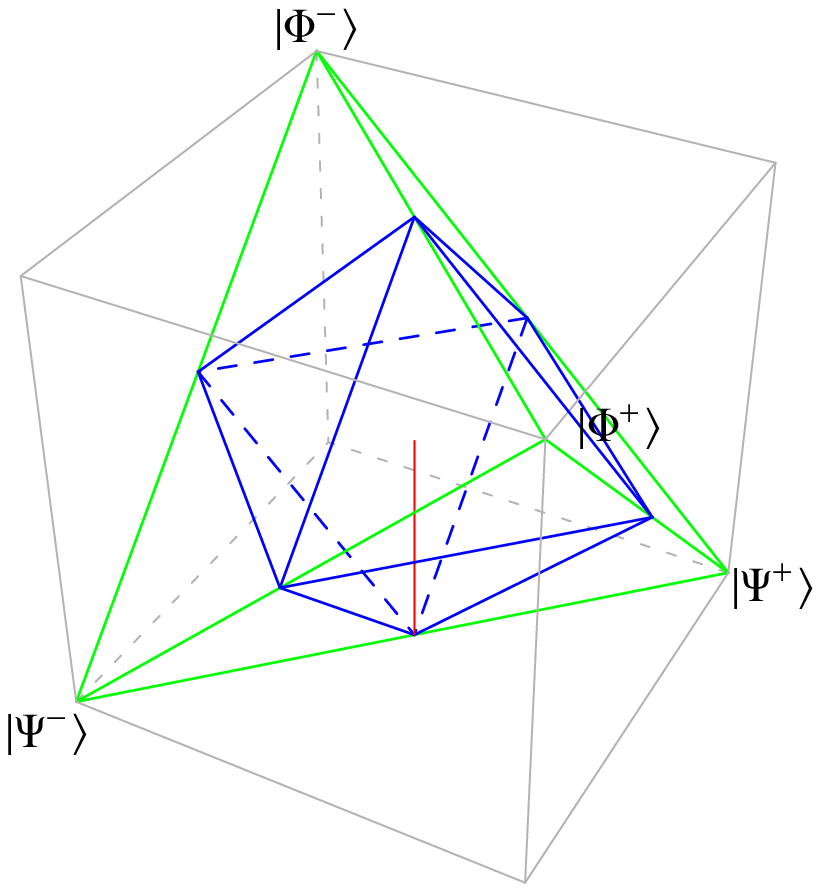}\label{fig.limit}}\qquad
\subfigure[]{\includegraphics[width=4.5cm,keepaspectratio=true]{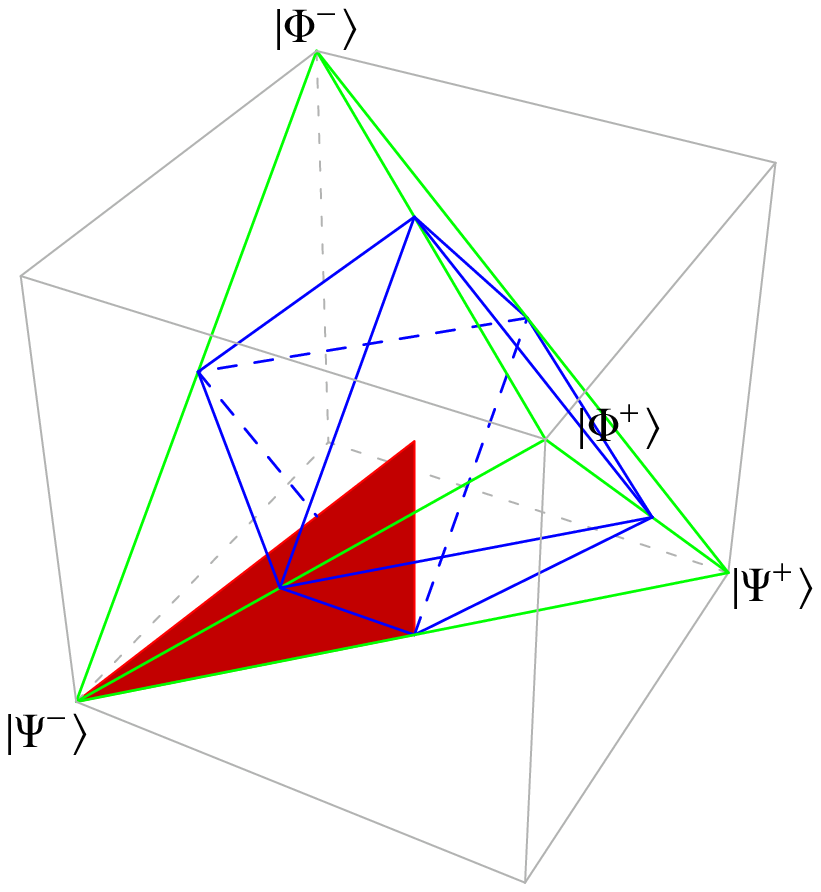}\label{fig.equiv}}%
\caption{Figure \subref{fig.limit} shows the asymptotic states of mode C in the spin geometry picture, given by Eq. \eqref{eq.limit-states}. In \subref{fig.equiv} all states which result in an equivalence of mode B and C are depicted. Note the symmetry of the picture of which we have picked out only one possibility.} 
\label{fig.limit_equiv}
\end{figure}

\textbf{Proposition.}\\
For a density matrix with initial local parameters $\vec m(0)$ and $\vec n(0)$ equal to zero and an initial correlation matrix given by
\begin{equation}\label{eq:initial-cond}
	c(0)= \begin{pmatrix}k_1 & 0 & k_2 \\0 & k_3 & 0 \\-k_2 & 0 & k_1\end{pmatrix}\;,
\end{equation}
the geometric picture of decoherence mode B and mode C coincide and the singular values depend only on the angle difference $\Delta$. The correlation vector of matrix \eqref{eq:initial-cond} is given by
$\vec c(0)=\begin{pmatrix}\sqrt{k_1^2+k_2^2}\;,& \sqrt{k_1^2+k_2^2}\;, & k_3\end{pmatrix}^T$
where the values $k_1$, $k_2$ and $k_3$ have to satisfy $2\sqrt{k_1^2+k_2^2}+k_3\leq 1$
in order to belong to the tetrahedron of possible states (see Fig. \ref{fig.equiv}).\\

\textbf{Proof.}\\
We consider states where the initial local parameters are set to zero because of their irrelevance for the spin geometry picture.

Two matrices $A$ and $B$ have the same eigenvalues $\lambda_i$ if they are similar $A=S B S^{-1}$ for an invertible matrix $S$. That means two conditions have to be satisfied:
\begin{itemize}
	\item $\tr A =\tr B=\sum_i \lambda_i$\;,
	\item $\det A =\det B=\prod_i \lambda_i$\;.
\end{itemize}
In our case we have to deal with the matrices $A^TA$ and $B^T B$ and find out when their eigenvalues are equal. In general the equality of the singular values of two matrices does not imply the equality of the eigenvalues of these matrices, e.g., in general
\begin{equation}
	A^T A=(S B S^{-1})^T (S B S^{-1})\neq S (B^T B) S^{-1}\;.
\end{equation}
The relation $A^T A= S (B^T B) S^{-1}$ is only valid for orthogonal transformation matrices $S$ where $S^{-1}=S^T$.
Let us assume that this condition is valid, then it is sufficient to consider only the conditions stated above\footnote{We have checked that $\tr (A^T A)$ and $\det (A^T A)$, in contrast to $\tr A$ and $\det A$, does not reveal new constraints on the coefficients.}. In the following we set $A:=c^B$ and $B:=c^C$.

There is only one point left which is important. The functions of $A$ depend only on the angle $\alpha$ which we have to substitute with the difference $\alpha-\beta$ in order to compare it with the functions of $B$. Then we calculate the difference of both solutions and determine the conditions for it to be zero and independent of the rotation angles.

We get for the trace condition 
\begin{equation}
\begin{split}
	\tr B-\tr A=&
	\left(1-e^{- \lambda t}\right)\sin\beta \cos (\alpha -\beta ) 
	\Bigl((c_{xz}+c_{zx}) \cos\alpha	+(c_{xx}-c_{zz}) \sin\alpha\Bigr) \;.
\end{split}
\end{equation}

This product is zero and independent of the single rotation angles only for
\begin{equation}\label{eq:condition1}
	c_{xx}=c_{zz}=k_1 \qquad\mbox{and}\qquad c_{xz}=-c_{zx}=k_2\;.
\end{equation}
All other entries of the correlation matrix are not fixed  by this condition and can be chosen arbitrarily. 

The condition of the determinant is given by 
\begin{equation}
\begin{split}
	\det B-\det A=&
	e^{-2 t \lambda } \left(1-e^{-\lambda t }\right) \sin\beta  \Bigl(k_1 \cos (\alpha -\beta )+k_3 \sin (\alpha -\beta )\Bigr) \\
	&\cdot\Bigl((c_{xy} c_{yz}+c_{yx} c_{zy}) \cos\alpha +(c_{xy} c_{yx}-c_{yz} c_{zy}) \sin\alpha\Bigr)\;,
\end{split}
\end{equation}
where we already used the condition stated in \eqref{eq:condition1}. This is equal to zero and independent of the rotation angles for
\begin{equation}\label{eq:condition2}
	c_{xy}=c_{yx}=c_{yz}=c_{zy}=0\;.
\end{equation}
Thus we have only one entry, namely $k_3$, in the initial correlation matrix left which is not determined by both conditions. $\Box$

Note that the invariants introduced in Refs.\cite{Makhlin,UshaDeviUmaPrabhuSudha} are connected with the problem considered here.

The states given by Eq. \eqref{eq:initial-cond} form planes (with proper restrictions of the coordinates) which contain the lines connecting the Bell states and the line of the Werner state (see Fig.~\ref{fig.modeB-1}) in the spin geometry picture.

\section{Summary and conclusion}\label{sect:summary}

We considered the master equation with a special type of dissipator which describes decoherence in a two qubit system. The generators of decoherence are chosen to be on the one hand projectors onto the eigenstates of the undisturbed Hamiltonian of the system (mode A) and on the other hand they are (bi-)local unitary rotations of these projectors (mode B and C). 

The general solutions of the master equation are presented with respect to the decomposition of a two qubit system in terms of joint Pauli matrices. We are interested in the time behavior of the correlation matrix which is studied for the case of the Bell singlet state. The information about entanglement and purity is encoded in the correlation vector which can be illustrated in the spin geometry picture. We show graphically the paths of the different decoherence modes  and discuss their behavior for the Bell singlet state (see Fig.~\ref{fig.modeB}).

For the special case of the Bell singlet state we find that mode B and C are equal and depend only on the difference of the rotation angles of the projection operators. This arises the question for which general initial states this is the case. We conclude that this happens for all states which are contained in the plane formed by the line connecting the Bell states and the line of the Werner states (see Fig.~\ref{fig.equiv}).

The asymptotic states of decoherence mode C (and therefore also for mode A and B) are found to give a line connecting the maximally mixed state at the origin of the picture with the equal mixture of two Bell states (see Fig.~\ref{fig.limit}).

The decoherence modes investigated in this paper exhibit very strong symmetry properties which is due to the fact that we choose projection operators as generators of the decoherence. For more general Lindblad operators the calculations start to get more involved but the conjecture is that for them the whole geometric state space of the tetrahedron can be occupied.

An open question related with the spin geometric picture is the influence of the local parameters on the geometric state space? We have some preliminary results for maximally entangled mixed states \cite{IshizakaHiroshima,MunroJamesWhiteKwiat} where the local parameters are not zero any more and the calculation of the singular values is more involved.

\begin{acknowledgments}

The author wants to thank Reinhold A. Bertlmann, Heide Narnhofer, Franz Embacher and Stefan Filipp for helpful discussions. 

This work has been supported financially by the Theodor-K\"{o}rner-Fond (F\"orderungspreis f\"ur Wissenschaft 2006), the University of Vienna (Forschungsstipendium) and the FWF project P 18943-N20 of the Austrian Science Foundation.

\end{acknowledgments}

\bibliographystyle{apsrev} 
\bibliography{/Users/kadu/Library/texmf/tex/latex/bibliography/bibliography-kadu}

\end{document}